\documentclass[preprint,prl,showpacs,showkeys]{revtex4-1}
\usepackage{amsfonts}
\usepackage{amsmath}
\usepackage{amssymb}
\usepackage{graphicx}
\usepackage{subfigure}

\begin{document}

\title{Anomalies in non-stoichiometric uranium dioxide induced by pseudo-phase transition of point defects}
\author{Hua Y. Geng}
\affiliation{National Key Laboratory of Shock Wave and Detonation
Physics, Institute of Fluid Physics, CAEP,
P.O.Box 919-102 Mianyang, Sichuan 621900, People's Republic of China}

\author{Hong X. Song}
\affiliation{National Key Laboratory of Shock Wave and Detonation
Physics, Institute of Fluid Physics, CAEP,
P.O.Box 919-102 Mianyang, Sichuan 621900, People's Republic of China}




\author{Q. Wu}
\affiliation{National Key Laboratory of Shock Wave and Detonation
Physics, Institute of Fluid Physics, CAEP,
P.O.Box 919-102 Mianyang, Sichuan 621900, People's Republic of China}


\begin{abstract}
A uniform distribution of point defects in an otherwise perfect crystallographic structure
usually describes a unique pseudo phase
of that state of a non-stoichiometric material.
With off-stoichiometric uranium dioxide as a prototype, we show that analogous to a conventional phase transition,
these pseudo phases also will transform from one state into another via changing the
predominant defect species when external conditions of pressure, temperature, or chemical composition
are varied.
This exotic transition is numerically observed along shock Hugoniots and
isothermal compression curves in UO$_{2}$ with first-principles calculations.
At low temperatures, it leads to anomalies (or quasi-discontinuities) in thermodynamic properties and electronic structures.
In particular, the anomaly is pronounced in both shock temperature and the
specific heat at constant pressure.
With increasing of the temperature, however, it transforms gradually to a smooth cross-over, and becomes
less discernible.
The underlying physical mechanism and characteristics of this type of transition are
encoded in the Gibbs free energy, and are elucidated clearly by analyzing the
correlation with the variation of defect populations as a function of pressure and temperature.
The opportunities and challenges for a possible experimental observation of this phase change are also discussed.

\end{abstract}

\volumeyear{year}
\volumenumber{number}
\issuenumber{number}
\eid{identifier}
\maketitle


Properties of states (or phases) of a matter and the transformation between them are
one of the central topics of modern condensed matter physics. What we mean by saying a phase
or a state of a matter refers to a region of space in a thermodynamic system that is chemically uniform, physically distinct,
and (often) mechanically separable. From a microscopic point of view, it usually constitutes
a sequence of static or dynamic distributions of particles, which possess specific symmetry or ordering, and
the physical properties are uniform, distinct, and time-independent in a statistical sense.
It is widely used as a classification of matters, for example, by differentiating
them according to the state
as solid, liquid, gas, or plasma, and subclassifying solid according to the crystallographic symmetry,
labeling magnetism by the ordering of magnetic moments, \emph{etc}.
When external conditions such as pressure, temperature, or chemical composition are varied,
a matter might change from one physical phase into another.
This transformation is often accompanied by a sudden modification in crystalline structure or ordering state, and can be recognized by an abrupt
change in thermodynamic properties.
On the other hand, an abrupt change (or singularity) in derivatives of free energy-- \emph{e.g.}, the specific volume, entropy, or specific heat--
usually signals a reorganization of the particles, and thus can be used as a helpful indicator of phase transitions.
Nevertheless, this is not always the case, especially in a complex system. In this paper, we will show
that anomalous or quasi-discontinuous jumps
take place in compression curves and other related thermodynamic quantities in non-stoichiometric
uranium dioxide. These abrupt changes, however, do not correspond to any known physical phase transitions from a conventional point of view.
Instead, we identify them as driven by a \emph{pseudo phase transition} that is a result of sharp switch of predominant defect species.

In principle, introduction of defects might modify or tune the physical properties of a material.
This modification, however, is primarily a result of accumulation, and reflects the magnitude of defect concentrations.
It is quite different from the abrupt variation in physical properties that arises from a symmetry or ordering change in a conventional phase transition.
Therefore, according to the conventional classification, a defective state of a material cannot be labeled
as a distinct physical phase against the perfect one.
Notwithstanding this, in some special situation the defect concentrations are strictly constrained
by, say, chemical
composition, and thus it has a possibility to exhibit a sudden change in predominant defect species
when thermodynamic conditions are varied, while keeping the underlying matrix structure almost unchanged.
In this sense, a uniform distribution of each type of defects would have some distinct physical properties,
and analogous to a conventional phase transition, might also lead to a jump in these properties when the predominant defect species is changed.
We therefore can identify each such kind of homogeneous distribution of defects
as a \emph{pseudo phase} according to the dominant species of defects to differentiate them.
As will be shown below, at low temperatures the variation in defect population during a change of pseudo phases is very
sharp, and strong anomalies or
quasi-discontinuities in physical properties due to pseudo phase transitions can be expected.

\begin{figure}
  \includegraphics*[width=3.5 in]{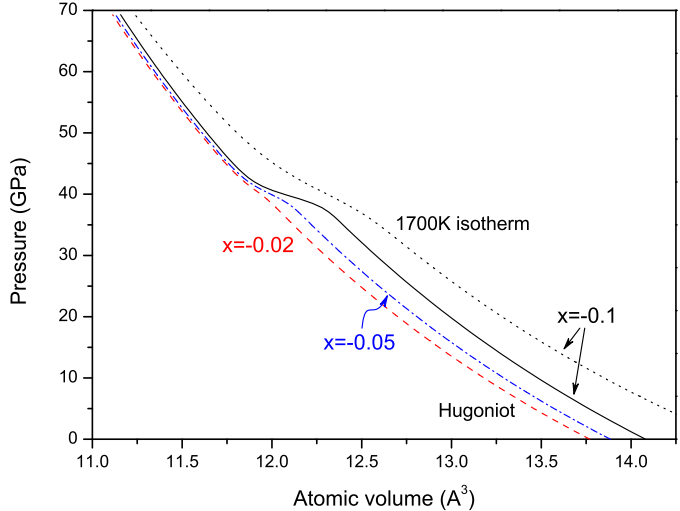}
  \caption{(Color online) Compression curves of UO$_{2+x}$ along the Hugoniot shocked from 500\,K and 0\,GPa as a function of the effective volume per atom at different
  stoichiometry deviations of $x$=-0.02, -0.05, and -0.1, respectively. An isotherm of
  $x$=-0.1 at 1700\,K is also plotted for comparison.
  }
  \label{fig:PV}
\end{figure}

Uranium dioxide (UO$_{2}$) is an important nuclear fuel. It is also a typical nonstoichiometric compound,\cite{catlow81}
in which the structure of the underlying matrix (in fluorite structure here) is stable over a wide range of pressure, and has
a great degree of stoichiometry deviation. It thus provides as a good prototype to
study the exotic effects of defects that manifest in thermodynamic properties.
For this purpose, we first computed the energetics of a series of defective UO$_{2}$ as a function
of pressure to extract the contribution of defects to formation free energies, from which
defect populations as a function of temperature and pressure were then calculated. 
After acquired these information, the Gibbs free energy as a function of chemical composition (or equivalently, the
deviation from stoichiometry) was constructed, as well as the equation of state (EOS) of the
non-stoichiometric system as a derived quantity.

In particular, the total energy calculations of the defective UO$_{2}$ system were carried out with the density functional theory using the PAW
pseudopotential\cite{blochl94,kresse99} plus plane-wave scheme as implemented in VASP code.\cite{vasp,kresse96} The LSDA+\emph{U} approach
was employed to treat the strongly correlated 5\emph{f} electronic orbitals that localized on uranium atoms, which
splits the \emph{f} band by an effective on-site Coulomb repulsion to form a Mott insulator
state.\cite{anisimov91} Details of the \emph{ab initio} calculations, as well as the establishment of non-stoichiometric
EOS, are referred to Ref.\cite{geng11}. Especially, in order to eliminate electronic meta-stable states that introduced
via strong correlations of 5\emph{f} orbitals, we adopted the quasi-annealing method\cite{geng2010} to lower
the electronic energy slowly by controlling the random disturbance from the ionic system which acts as a kind of
heat bath. Defect populations were evaluated using point defect model\cite{lidiard66,matzke87,geng08} and presumed a physical
condition in the closed regime, which corresponds to the interior of a bulk material and
thus no exchange of particles with the exterior can occur.

\begin{figure}
  \includegraphics*[width=3.0 in]{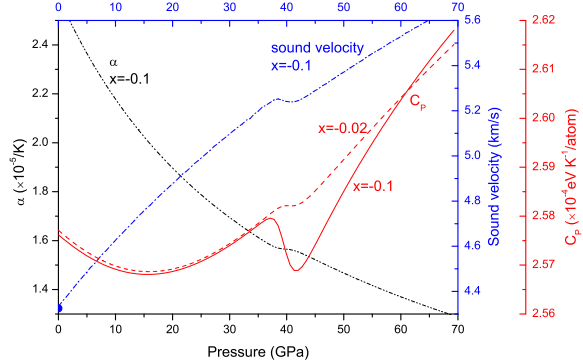}
  \caption{(Color online) Anomalies in the thermal expansivity $\alpha$,
  isothermal bulk sound velocity, and specific heat at constant pressure $C_{P}$ along the
  Hugoniot that shocked from 500\,K and 0\,GPa in non-stoichiometric UO$_{2}$. The solid circle (at the bottom left corner) marks the experimental bulk
  sound velocity measured at ambient conditions
  in perfect UO$_{2}$.
  }
  \label{fig:CpP}
\end{figure}

With these calculations, an interesting phenomenon was observed in the
hypo-stoichiometric regime ($x<0$) of UO$_{2+x}$.
Figure \ref{fig:PV} shows the Hugoniots that shocked from an initial condition of 500\,K and 0\,GPa
for $x=-0.02$, $-0.05$, and $-0.1$, respectively,
by comparing with an isotherm of $x=-0.1$ at 1700\,K.
A volume collapse was observed at about 39\,GPa.
Note that in calculations we fixed the underlying matrix in a fluorite structure all the time,
therefore this volume collapse doesnot correspond to any structural phase transformation.
Nevertheless, this exotic behavior in the compression curve is very similar to what
happens in a first-order phase transition.

It is worthwhile to point out that the magnitude of this volume change is proportional to
the size of stoichiometry deviation $x$, which is more pronounced for $x=-0.1$ than that
for $x=-0.02$. It is a direct consequence of accumulation effect of defect populations,
and is quite different from a first-order phase transition in which the dependence of the volume collapse
on stoichiometry deviation should be weak. We also observed that the sharp volume change
becomes smooth, and even less distinguishable, at higher temperatures, as the isotherm at 1700\,K in Fig.\ref{fig:PV} illustrates,
which reminds us that it is not a conventional phase transition.

Furthermore, for non-stoichiometric UO$_{2}$, it is not only compression curve that exhibits quasi-discontinuity,
other thermodynamic quantities also manifest similar anomalies.
Figure \ref{fig:CpP} plots the calculated
thermal expansivity $\alpha$, specific heat at constant pressure $C_{P}$, and isothermal bulk sound velocity
that defined as $\rho C^2=-V\left({\partial P}/{\partial V}\right)_{T}$ along the same shock
Hugoniots as in Fig.\ref{fig:PV}.
Here $\rho$, $P$, $V$, and $T$ denote the mass density,
pressure, specific volume, and temperature, respectively.
An experimental data of bulk sound velocity of perfect UO$_{2}$ measured at ambient conditions\cite{fritz76}
is also drawn for comparison.
It can be seen that all of these thermodynamic quantities show a kink at the same pressure of about 39\,GPa.
It is interesting to note that the anomaly in $C_{P}$ is much more striking than that of others,
and is still discernible even for $x=-0.02$. By comparison, the expansivity $\alpha$ already shows a nearly smooth crossover for $x=-0.1$,
whereas at the same $x$ the quasi-discontinuous change in sound velocity is just perceptible.
The occurrence of quasi-discontinuity in both the first and second derivatives of
the free energy at the transition point implies that although this transformation
belongs to neither a first nor a second order phase transition, it has
some features that are analogous to both of them.

\begin{figure}
  \includegraphics*[width=3.5 in]{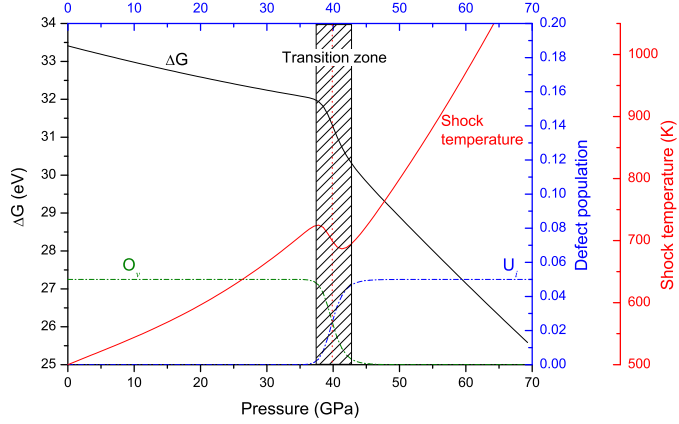}
  \caption{(Color online) Variations of the shock temperature, defect populations, and the difference of Gibbs free
  energy with respect to the perfect UO$_{2}$ when across the transition zone, respectively.}
  \label{fig:GP}
\end{figure}

In fact, the physical mechanism of this transformation and the associated anomalies are completely
encoded in the construction of
the Gibbs free energy of a non-stoichiometric system, which reads\cite{geng11}
\begin{equation}
  G(T,P,x)=G^{0}(T,P) + \Delta G(T,P,n),
  \label{eq:gibbs}
\end{equation}
where $G^{0}$ is the Gibbs free energy of the perfect system and $\Delta G$ is that contributed
by defects. The defect populations $n$ are constrained by the composition equation of
\begin{equation}
  f(n_{1},n_{2},\ldots,n_{i})=x.
\end{equation}
To the first-order approximation, $G$ can be written as
\begin{equation}
  G\approx G^{0}+\sum_{i}{\delta G_{i}}n_{i}.
  \label{eq:Gtot}
\end{equation}
Since $\delta G_{i}$ differs from each other, the Gibbs free energy thus could have a cross-over when
the predominant defect switches from one type into another under a variation of external conditions
of temperature, pressure, or chemical composition. Furthermore, since distribution of point defects satisfies
Boltzmann statistics, the population variation is in proportion to $\exp(-\delta G/k_{B}T)$.
Therefore the crossover is smooth at high temperatures,
but becomes sharp at low enough temperature,
which leads to a discontinuity or strong anomaly in thermodynamic properties.
It is necessary to note that since the Gibbs free energy in Eq.(\ref{eq:Gtot}) is
analytic, all of its derivatives must be continuous at finite temperatures in nature, even though
some of them might appear to be \emph{quasi-discontinuous}.

\begin{figure}
  \includegraphics*[width=3.0 in]{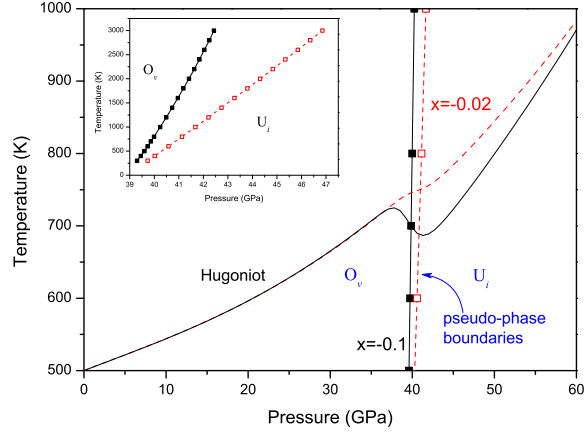}
  \caption{(Color online) Pseudo-phase diagram of hypo-stoichiometric UO$_{2}$ on $T$-$P$ plane, together
  with the induced anomaly in shock temperature.
  Solid lines denote that for $x$=-0.1 and dashed lines for $x$=-0.02. Inset provides an
  overall view of the boundaries that separate the dominant defects of oxygen vacancy and uranium
  interstitial for $x$=-0.02 (open square) and -0.1 (filled square) respectively.
  }
  \label{fig:TP}
\end{figure}

Figure \ref{fig:GP} shows the variation of shock temperature of $x=-0.1$, in which
the kink is much more appreciable than in $P$-$V$ curve. Also, the difference of the Gibbs
free energy with respect to the perfect UO$_{2}$ (\emph{i.e.} $\Delta G$) and the defect populations
as a function of shock pressure are plotted.
It can be seen that the
kink in temperature corresponds exactly to the point of switch of predominant defect species.
At low pressure it is oxygen vacancy ($\mathrm{O}_{v}$) that dominates.
But at a pressure higher than 39\,GPa,
uranium interstitial ($\mathrm{U}_{i}$) starts to take over, and $\mathrm{O}_{v}$ is suppressed
completely.
This switch also reflects in the Gibbs free energy as a function of
shock pressure, and $\Delta G$ (\emph{i.e.}, that contributed by defects) displays a clear cross-over from
one end to the another when crossing the transition zone.
These demonstrate that the anomalies are results of a pseudo-phase transition in non-stoichiometric UO$_{2}$ driven by compression,
as above discussions implied.

If we define the locus of the intersection points of defect populations
(usually at the transition zone center) as a pseudo-phase boundary, a pseudo-phase diagram
can be constructed, which divides the dominant regime of different defect species. As shown
in Fig.\ref{fig:TP}, for hypo-stoichiometric UO$_{2}$, O$_{v}$ dominates the low pressure
region and $\mathrm{U}_{i}$ is the main defective component at high pressures.
Also, we found that decreasing $x$ increases the transition pressure slightly. The change
is about several giga-pascal at high temperatures, but less than 1\,GPa when below 500\,K, as
the inset of Fig.\ref{fig:TP} shows.
Overall, the temperature dependence of the
pseudo-phase boundary is weak, reflecting the fact that this transition is mainly driven by compression.

\begin{figure}
  \includegraphics*[width=3.0 in]{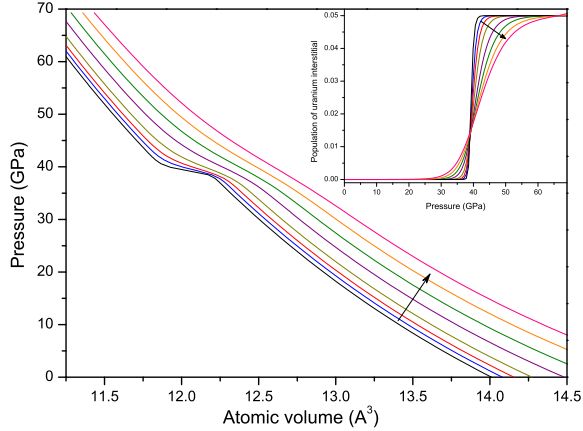}
  \caption{(Color online) Isothermal compression curves for $x$=-0.1 at temperatures of 300, 500, 700,
  1000, 1500, 2000, 2500, and 3000\,K, respectively.
  Inset: defect population of uranium interstitial along the isotherms.
  Arrows indicate the direction of temperature increase.
  }
  \label{fig:isoth}
\end{figure}

It is worthwhile to point out that the width of the transition zone increases with temperature. This
is a direct reflection of the dependence of defect populations on temperature.
At a high enough temperature, the transition zone will become so wide that
the cross-over behavior is almost indiscernible, and then the quasi-discontinuities (or kinks)
in thermodynamic quantities disappear. On the other hand, it should be noted that theoretically a pressure-induced pseudo-phase transition at absolute zero Kelvin
is possible, which, however, requires a large zero point
motion of particles to facilitate ionic diffusion so that to make the switch of defect species possible.
In this extreme case, the width of the transition zone becomes zero, and the Gibbs
function has a sharp kink so that all of its derivatives become truly discontinuous at the transition point.
It looks more like a first-order transition. The difference is that all derivatives
of the free energy higher than the first order
are divergent at the transition point for a first-order phase transition, while they
are finite for a zero Kelvin pseudo-phase transition.
This difference comes from the fact that for a defect type switch transition, the derivatives
are carried out with defect populations being fixed, which is well defined even at the transition
point. But it is not that case for a firs-order phase transition, and only left (or right) limit can be defined.
Figure \ref{fig:isoth} illustrates the characteristic of
evolution of the crossover from being quasi-discontinuous at low temperatures to a smooth bridging
at high temperatures. The variation of the population of $\mathrm{U}_{i}$ as shown in the inset of Fig.\ref{fig:isoth} as
a function of pressure and temperature, which exhibits a sharp jump at low $T$ but becomes a smooth
crossover at high $T$, explains perfectly all anomalous behavior observed in the Gibbs free energy and
its derivatives.

In above discussion, we did not consider the possibility of a phase segregation, or equivalently, the stability
of hypo-stoichiometric UO$_{2}$. Experiments indicated that the stable range of $x$ at low
temperature is very narrow, especially in the $x<0$ side.\cite{higgs07,lewis04} Nevertheless, a meta-stable UO$_{2+x}$ with
$x=-0.02$ had been successfully synthesized at room temperature in laboratory.\cite{kapshukov90}
Although above discussions showed that the anomalies for this chemical composition
is weak in the specific volume or shock temperature, a kink
in the $C_{P}$ curve is still evident, which implies that the pseudo-phase transition
can be detected by a highly accurate measurement of $C_{P}$ in UO$_{1.98}$.
Furthermore, it is also possible that a meta-stable phase
with $x=-0.05$ can be synthesized at high temperatures. This
stoichiometry deviation would provide more noticeable
anomalies, and thus is much easier to detect it.
In addition, according to the phase diagram of UO$_{2+x}$, when temperature is higher than 1000\,K, a composition
of $x=-0.1$ might become possible. At this temperature range the crossover behavior should have been
smeared drastically, but our calculations indicate that some of the thermodynamic properties,
especially $C_{P}$, still exhibit a strong anomaly that can be detected by experiment.

Another challenging issue for experiments to detect a pseudo phase transition
is the possible interference coming from structural transition of the underlying matrix.
Our calculations suggested that the pseudo transition in UO$_{2+x}$ might take place at about 39\,GPa
at room temperature. For perfect UO$_{2}$, however, compression also
leads to a phase transition from fluorite structure to cotunnite phase at $\sim$42\,GPa.\cite{idiri04,geng07}
Considering that introduction of defects should weaken the stability of the
matrix, the actual transition pressure of the underlying matrix
for off-stoichiometric compound thus might be lowered to overlap with that of the pseudo phase transition,
and make the relevant experiments very difficult.
Nonetheless, what we considered
here is only neutral defects. Since UO$_{2}$ is an insulator within this pressure range, the defects
can be charged. This would modify the energetics of
all defects.\cite{crocombette11} It therefore still has the possibility that the $\mathrm{O}_{v}$-$\mathrm{U}_{i}$ switch
occurs at a much lower pressure.
To make a reliable assessment of this,
however, requires a comprehensive evaluation of the energetics of charged
defects in UO$_{2}$, which is beyond the scope of this paper.
In spite of that, our analyses definitely demonstrated that a pseudo-phase transition is practically
possible in non-stoichiometric compounds.

According to the defect-type switching mechanism, a pseudo phase transition driven by temperature
or chemical composition rather than by compression is also possible. Similar anomalies
can be expected for all of these transitions. In fact, the anomalies in thermal expansion
coefficient and bulk modulus due to defect-type switch might have been observed experimentally in Fe$_{1-x}$O.
A large discontinuous jump in thermal expansivity and bulk modulus were observed when the chemical composition
across the point of Fe$_{0.98}$O.\cite{zhang05,zhang00} The underlying physical mechanism was not clearly understood at that time, and
it was proposed might be related to defect clustering as $x$ increased. It becomes clear now
that this picture is perfectly in line with the pseudo phase transition just discussed above
if we treat the defect clusters as independent species,
\emph{i.e.}, it is the switch of defect species that leads to the observed quasi-discontinuous
jump in thermodynamic properties. Furthermore, we noticed that the transition in Fe$_{1-x}$O is very similar to
what would happen in a hyper-stoichiometric UO$_{2}$, where the stoichiometry deviation drives
a transition from point oxygen interstitial to COT-$o$ cluster, as the pseudo phase diagram
of Fig.3 in Ref.\cite{geng08c} implied.
However, it should be noted that it is difficult to control the stoichiometry deviation continuously,
thus it is still very important to study by experiment the pseudo phase transitions that driven by pressure,
which can provide a more explicit and clear-cut physical picture.

In summary, we presented in this paper that a pseudo-phase transition can occur in non-stoichiometric
compounds (and alloys as well) which arises from a rapid switch of predominant defect species
driven by external conditions of pressure, chemical composition, or temperature, and so on.
At high temperatures this transition is a smooth crossover, whereas it becomes quasi-discontinuous
at low temperatures and has features similar to a first-order phase transition. A real discontinuity
might be possible at 0\,K, but which requires a large zero point motion of ions
to facilitate particle diffusion, therefore is only possible in compounds of light elements.
Being analogous to a physical phase transition, anomalies also present in thermodynamic properties
at the transition point of a pseudo phase transformation. With hypo-stoichiometric UO$_{2}$ as a prototype, we demonstrated that such
kind of anomalies are evident in both shock Hugoniots and isothermal compression curves using
\emph{ab initio} calculations. The transition mechanism is clearly elucidated
by analysis of the Gibbs free energy and evolution of defect populations as a function
of pressure, stoichiometry deviation, and temperature. We also discussed the possibility to observe
this transition in non-stoichiometric UO$_{2}$ by experiment, as well as the accompanied challenges.
It is worthwhile to point out that since electronic structure and magnetic property usually
are governed (or tuned) by defect species, a pseudo phase transition
thus indicates a rapid change of these properties (in addition to the thermodynamic properties
that are direct derivatives of the Gibbs free energy).
In this regard, the effect of pseudo phase transition of defects might have potential impacts on
electronic applications, for example, in
pressure-sensitive switchers that change the electronic state of a device from one state into
another by slightly adjusting the stress state.

\begin{acknowledgments}
Support from the Fund of National Key Laboratory of Shock Wave and Detonation Physics of China
(under Grant No. 9140C6703031004) is acknowledged.
\end{acknowledgments}



\begin{thebibliography}{99}
\bibitem{catlow81} C. R. A. Catlow, in \emph{Nonstoichiometric oxides}, edited by O. T. S{\o}rensen (Academic, New York, 1981).
\bibitem{blochl94} P. E. Bl{\"o}chl, Phys. Rev. B \textbf{50}, 17953 (1994).
\bibitem{kresse99} G. Kresse and D. Joubert, Phys. Rev. B \textbf{59}, 1758 (1999).
\bibitem{vasp} G. Kresse and J. Furthm{\"u}ller, Comput. Mater. Sci. \textbf{6}, 15 (1996).
\bibitem{kresse96} G. Kresse and J. Furthm{\"u}ller, Phys. Rev. B \textbf{54}, 11169 (1996).
\bibitem{anisimov91} V. I. Anisimov, J. Zaanen, and O. K. Andersen, Phys. Rev. B \textbf{44}, 943 (1991).
\bibitem{geng11} H. Y. Geng, H. X. Song, K. Jin, S. K. Xiang, and Q. Wu, Phys. Rev. B \textbf{84}, 174115 (2011).
\bibitem{geng2010} H. Y. Geng, Y. Chen, Y. Kaneta, M. Kinoshita, and Q. Wu, Phys. Rev. B \textbf{82}, 094106 (2010).
\bibitem{lidiard66} A. B. Lidiard, J. Nucl. Mater. \textbf{19}, 106 (1966).
\bibitem{matzke87} Hj. Matzke, J. Chem. Soc., Faraday Trans. {2} \textbf{83}, 1121 (1987).
\bibitem{geng08}  H. Y. Geng, Y. Chen, Y. Kaneta, M. Iwasawa, T. Ohnuma, and M. Kinoshita, Phys. Rev. B \textbf{77}, 104120 (2008).
\bibitem{fritz76} I. J. Fritz, J. Appl. Phys. \textbf{47}, 4353 (1976).
\bibitem{higgs07} J. D. Higgs, W. T. Thompson, B. J. Lewis, and S. C. Vogel, J. Nucl. Mater. \textbf{366}, 297 (2007).
\bibitem{lewis04} B. J. Lewis, W. T. Thompson, F. Akbari, D. M. Thompson, C. Thurgood, and J. Higgs,
J. Nucl. Mater. \textbf{328}, 180 (2004).
\bibitem{kapshukov90} I. I. Kapshukov, N. V. Lyalyushkin, L. V. Sudakov, A. S. Bevz, and O. V. Skiba,
J. Radioanal. Nucl. Chem. \textbf{143}, 213 (1990).
\bibitem{idiri04} M. Idiri, T. Le Bihan, S. Heathman, and J. Rebizant, Phys. Rev. B \textbf{70}, 014113 (2004).
\bibitem{geng07}  H. Y. Geng, Y. Chen, Y. Kaneta, and M. Kinoshita, Phys. Rev. B \textbf{75}, 054111 (2007).
\bibitem{crocombette11} J. P. Crocombette, D. Torumba, and A. Chartier, Phys. Rev. B \textbf{83}, 184107 (2011).
\bibitem{zhang05} J. Zhang and Y. Zhao, Phys. Chem. Minerals \textbf{32}, 241 (2005).
\bibitem{zhang00} J. Zhang, Phys. Rev. Lett. \textbf{84}, 507 (2000).
\bibitem{geng08c}  H. Y. Geng, Y. Chen, Y. Kaneta, and M. Kinoshita, Appl. Phys. Lett. \textbf{93}, 201903 (2008).



\end{thebibliography}
\end{document}